\documentclass[pra, reprint, twocolumn, amsmath,amsfonts,amssymb]{revtex4-1}
\usepackage{graphicx}
\usepackage{amsmath,amsfonts,amssymb}
\usepackage[colorlinks=true,linkcolor=blue]{hyperref}
\usepackage{color}
\usepackage{bm}
\usepackage{threeparttable}
\usepackage[mathscr]{eucal}
\usepackage{picture}

\usepackage{tablefootnote}
\usepackage{tikz}
\usetikzlibrary{quantikz}

\usepackage{color}
\usepackage{bm}
\definecolor{grey}{rgb}{0.7,0.7,0.7}
\newcommand{\black}{\color{black}{}}

\usepackage{fancyvrb}
\usepackage{ulem}

\usepackage[T1]{fontenc}
\usepackage{braket}
\allowdisplaybreaks
\begin{document}
\title{Multi-state quantum simulations via model-space quantum imaginary time evolution}
\author{Takashi Tsuchimochi}
\email{tsuchimochi@gmail.com}
\affiliation{Graduate School of System Informatics, Kobe University, 1-1 Rokkodai-cho, Nada-ku, Kobe, Hyogo 657-8501 Japan}
\affiliation{Japan Science and Technology Agency (JST), Precursory Research for Embryonic Science and Technology (PRESTO), 4-1-8 Honcho Kawaguchi, Saitama 332-0012 Japan}
\author{Yoohee Ryo}
\affiliation{Graduate School of Science, Technology, and Innovation, Kobe University, 1-1 Rokkodai-cho, Nada-ku, Kobe, Hyogo 657-8501 Japan}
\author{Siu Chung Tsang}
\affiliation{Graduate School of System Informatics, Kobe University, 1-1 Rokkodai-cho, Nada-ku, Kobe, Hyogo 657-8501 Japan}

\author{Seiichiro L. Ten-no}
\affiliation{Graduate School of System Informatics, Kobe University, 1-1 Rokkodai-cho, Nada-ku, Kobe, Hyogo 657-8501 Japan}

\begin{abstract}
We introduce the framework of model space into quantum imaginary time evolution (QITE) to enable stable estimation of ground and excited states using a quantum computer. Model-space QITE (MSQITE) propagates a model space to the exact one by retaining its orthogonality, and hence is able to describe multiple states simultaneously. The quantum Lanczos (QLanczos) algorithm is extended to MSQITE to accelerate the convergence. The present scheme is found to outperform both the standard QLanczos and the recently proposed folded-spectrum QITE in simulating excited states. Moreover, we demonstrate that spin contamination can be effectively removed by shifting the imaginary time propagator, and thus excited states with a particular spin quantum number are efficiently captured without falling into the different spin states that have lower energies. We also investigate how different levels of the unitary approximation employed in MSQITE can affect the results. 
The effectiveness of the algorithm over QITE is demonstrated by noise simulations for the H$_4$ model system.
\end{abstract}
\maketitle

\section*{Introduction}
Variational quantum algorithms\cite{Moll18,Cerezo21,Farhi14} are expected to play a key role on noisy intermediate-scale quantum (NISQ) devices\cite{Preskill18}. Especially, variational quantum eigensolver (VQE)\cite{Peruzzo14, Wang19, McClean16, Wecker15, McClean16, OMalley16,Kandala17} has attracted much attention for its application to quantum chemistry where quantum entanglement is essential\cite{Cao19, McArdle20}. The scope of VQE has extended from ground state simulations of molecular systems\cite{OMalley16, McClean16, Grimsley19} to condensed matters\cite{Cerasoli20,Fan21,Yoshioka22} and excited states\cite{McClean17, Higgott19, Parrish19, Nakanishi19,Zhang21A, Xie22}.

Building on the concept of imaginary time evolution (ITE), which aims to drive an arbitrary initial state to the exact ground state, several variants of quantum algorithms have recently been developed.
McArdle and co-workers proposed variational ITE (VITE)\cite{McArdle19}, which employs a fixed ansatz and then determines the optimal parameters using McLachlan’s variational principle. Therefore, VITE can be regarded as an optimizer not only for variational algorithms such as VQE\cite{Benedetti21}, but also for algorithms that employ pre-processed non-Hermitian Hamiltonians such as transcorrelated methods\cite{transcorrelated1, transcorrelated2, transcorrelated3}.   Probabilistic ITE (PITE) exploits measurements to perform the non-unitary operation of ITE on quantum devices probabilistically\cite{Lin21, Liu21,kosugi22}. This can be achieved by introducing one ancilla qubit and embedding the ITE operator acting on an $N$-qubit system in an ($N+1$)-qubit unitary gate\cite{Lin21}. In principle, ITE can be exactly performed on the  $N$-qubit system by accepting (or discarding) the resultant state if the ancilla qubit is measured to be $|0\rangle$ (or $|1\rangle$). However, the exact PITE generally requires the singular-value-decomposition of the $N$-qubit Hamiltonian, which hinders its practical applications in chemistry. Furtheremore, the success probability decays exponentially with the number of imaginary time steps and the system size $N$. As such, there have been several proposals to circumvent these problems\cite{Liu21, kosugi22}.

Yet another scheme, Quantum ITE (QITE)\cite{Motta20}, approximates the non-unitary short evolution of ITE by a unitary evolution that is determined by solving a set of linear equations. Therefore, it circumvents the high-dimensional noisy optimizations in variational algorithms, while driving a quantum state towards the ground state at each evolution step. The promise of QITE has been demonstrated experimentally\cite{Motta20, Yeter-Aydeniz20, Yeter-Aydeniz21}, and numerous authors have extended the algorithm\cite{Gomes20, Gomes21, Sun21, Huang22, Amaro22, Tsuchimochi22B, Jouzdani22}. In our own recent study, a modified equation for the unitary approximation was presented, which enables faster convergence of QITE, thereby reducing the overall quantum resources. 

Although QITE is a powerful tool for determining the ground state, there have been fewer developments that aim for obtaining excited states, especially when compared to variational algorithms that have seen a wide variety of developments\cite{McClean17, Santagati18, Colless18,Higgott19, Parrish19, Nakanishi19, Ollitrault20, Zhang21A,  Tkachenko22, Heya23, Tsuchimochi23A,  Yoshikura23}. The reason for this is perhaps that quantum Lanczos diagonalization (QLanczos) is expected to find reasonable excited states by increasing the size of the Krylov subspace\cite{Motta20, Yeter-Aydeniz20}. However, our recent study showed that the component of excited states encoded in the initial state vanishes with imaginary time $\beta$ at an exponential rate in general, and is lost in the numerical noise that is caused by the strong linear dependence of the chosen Krylov subspace\cite{Tsuchimochi22B}. This is particularly true if the excited states are separated from the ground state by large energy gaps, i.e., higher energy eigenstates.

Historically, there have been broad interests in obtaining excited states from classical ITE\cite{Ohtsuka10, Booth12, Tenno13, Blunt15, Tenno17}, and we can gain many insights from them. For instance, to retain the excited state signature throughout the QITE simulation, we followed the work of Booth and Chan\cite{Booth12} and adopted the folded-spectrum propagator $e^{-\beta^2 (\hat H - \omega)^2}$ in Ref.~[\onlinecite{Tsuchimochi22B}], an approach coined FSQITE. It was shown that FSQITE can in principle yield the desired excited states, and its convergence rate can be drastically accelerated with QLanczos. Nevertheless, FSQITE requires to estimate the target energy $\omega$ in advance and to treat the Hamiltonian squared $\hat H^2$, which can be quite challenging in general. 

In this work, we develop the model-space QITE (MSQITE) algorithm to deliver stable and accurate solutions for excited states. MSQITE evolves an orthogonal model space to the complete subspace by ITE, simulating multiple states simultaneously. It also improves the behavior and accuracy for the ground states of strongly correlated systems, by directly incorporating important configurations. Because the method has many similarities to QITE and FSQITE, it can be also easily combined with QLanczos.

Furthermore, we present a scheme to deal with spin contamination in MSQITE. The spin quantum number is an essential quantity that characterizes a non-relativistic electronic state. Preserving spin symmetry is important but is more challenging in quantum simulations\cite{Liu19, Gard20, Seki20, Yen19, Tsuchimochi20,Tsuchimochi22} than conserving other symmetries such as point-group symmetry that can be usually constrained by removing the qubits from the simulation\cite{Bravyi17,Setia20}. It should be easily imagined the problem of spin contamination is exacerbated in excited state calculations, because excited states often exhibit more complicated electronic structures than the ground state and thus are prone to spin contamination. In the following, we provide a way to circumvent this difficulty.

As will be seen, there are two different flavors of MSQITE; one uses the same unitary for all states in the model space, and the other employs different unitaries for different states. We investigate how such unitary approximations in the MSQITE algorithm can affect its representability and accuracy, and report difficulties with the former approach. 

Since the quantum circuits of both QITE and MSQITE necessarily elongate with imaginary time, their applicability on NISQ computers might be severely limited due to quantum noise. However, it is expected that MSQITE could potentially outperform QITE in many respects, even on NISQ computers, providing not only excited states but also faster convergence to the ground state. We demonstrate the efficacy of MSQITE through noisy simulations using a simple error mitigation protocol.


\section*{Results}\label{sec:Theory}
 
\subsection*{Model-space QITE} \label{sec:MS}
In MSQITE, one prepares an orthogonal subspace that consists of zeroth-order ground and excited states, $\{|\Phi_I\rangle; I = 0, \cdots, n_{\rm states} - 1\}$, and evolves the entire subspace by the propagator $e^{-\beta \hat H}$ (which is Trotterized by a short time step $\Delta\beta$). It is important to note that the imaginary time evolution makes the basis states nonorthogonal, $\langle \Phi_I|e^{-2\Delta\beta\hat H}|\Phi_J\rangle\ne 0$, and therefore the orthonormalization of the subspace is necessary. Hence, in MSQITE, we consider the following unitary approximation on the $\ell$th step:
\begin{align}
	|\Phi_I^{(\ell +1)}\rangle = \sum_{J} d_{IJ}  e^{-\Delta \beta (\hat H - E_J)}|\Phi_J^{(\ell)}\rangle \approx e^{-i \Delta\beta \hat A} |\Phi_I^{(\ell)}\rangle\label{eq:MSQITE}
\end{align}
where $|\Phi_I^{(\ell)}\rangle$ are the $I$th state at the $\ell$th time step, and $\hat A$ is a Hermitian operator parameterized by the coefficients {\bf a},
\begin{align}
	\hat A = \sum_{\mu} a_\mu \hat \sigma_\mu \label{eq:A}
\end{align}
with Pauli strings $\hat \sigma_\mu$, which are appropriately chosen\cite{Motta20, Gomes20, Tsuchimochi22B}. In Eq.~(\ref{eq:MSQITE}), we have intentionally introduced the energy shift $E_J = \langle \Phi_J^{(\ell)} | \hat H | \Phi_J^{(\ell)}\rangle$ for convenience. 
The transformation matrix {\bf d} is also introduced to ensure the orthonormality of the time-evolved model space $\{|\Phi_I^{(\ell +1)}\rangle\}$. 

This transformation matrix can be defined in infinitely different ways; however, we require ${\bf d} \rightarrow {\bf I}$ (identity matrix) as $\Delta\beta\rightarrow 0$, because this would allow us to correctly obtain $|\Phi^{(\ell+1)}_I\rangle\equiv |\Phi_I^{(\ell)}\rangle$. We also wish the change in each state to be minimum at each time step, to be able to ``follow'' the $I$th state between the time steps in order for Eq.~(\ref{eq:MSQITE}) to be a meaningful approximation. To this end, we employ the L\"owdin symmetric orthonormalization\cite{Lowdin50, Szabo}. Remarkably, the so-obtained {\bf d} is the one that minimizes the distance in the Hilbert space, ${\bf d} = \arg\min_{{\bf d}}\sum_I \| |\Phi_I^{(\ell+1)}\rangle - e^{-\Delta\beta (\hat H-E_I)}|\Phi_I^{(\ell)}\rangle \|^2$\cite{Carlson57, Mayer02}. In other words, the property of $|\Phi_I^{(\ell)}\rangle$ is maximally preserved in $|\Phi_I^{(\ell+1)}\rangle$ on average, and therefore it is expected that different states do not mix strongly. In particular, when the energy shift $E_J$ is introduced, {\bf d} is diagonal dominant with all the diagonal elements being equal to one. 
In the Methods section, we have detailed the L\"owdin symmetric orthonormalization procedure in MSQITE and discussed other possibilities for the definition of {\bf d}.

In MSQITE, $\lim_{\ell \rightarrow \infty}|\Phi_I^{(\ell)}\rangle$ may {\it not} be the exact ground and excited states. Instead, we retain them as a model space basis and express the physical states $|\psi_I\rangle$ as a linear combination of these states, $|\psi_I\rangle =\lim_{\ell\rightarrow \infty} \sum_K c_{KI} |\Phi_K^{(\ell)}\rangle$. This corresponds to solving the eigenvalue problem
\begin{align}
	{\bf H}^{(\ell)}{\bf c} = {\bf S}^{(\ell)}{\bf c} {\bm {\mathcal E}} \label{eq:HcScE}
\end{align}
where
\begin{align}
	H^{(\ell)}_{IJ} &= \langle \Phi_I^{(\ell)}| \hat H|\Phi_J^{(\ell)}\rangle\label{eq:HIJ}\\
	S^{(\ell)}_{IJ} &= \langle \Phi_I^{(\ell)}| \Phi_J^{(\ell)}\rangle	
\end{align}
and ${\bm {\mathcal E}}$ contains the ground and excited state energies in the diagonal. The eigenvalues become the exact energies if the entire model space is propagated appropriately. 

\begin{figure*}
	\includegraphics[width = 170mm]{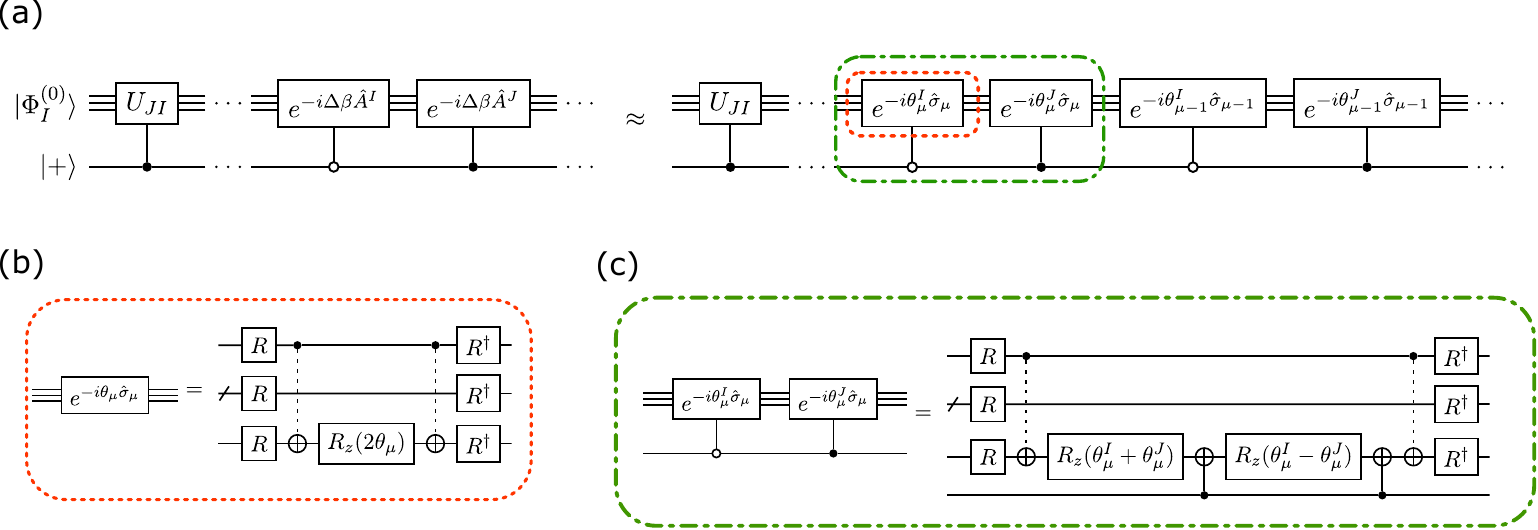}
	\caption{Quantum circuit for obtaining matrix elements of state-specific MSQITE using the Hadamard test with an ancilla $|+\rangle$. (a) $U_{JI}$ transforms $|\Phi_I^{(0)}\rangle$ to $|\Phi_J^{(0)}\rangle$. The unitary gates $e^{-i \Delta\beta \hat A^I}$ and $e^{-i \Delta\beta \hat A^J}$ are each decomposed into Pauli rotations. (b) The uncontrolled version of Pauli rotation is typically implemented by using one-qubit unitary gates ($R$), and a sequence of CNOT gates, which is abbreviated by the CNOT gate with a dotted line. Together with $R_z$, they perform $e^{-i\theta_\mu\hat \sigma_\mu}$. (c) Two different controlled Pauli rotations by $\theta_\mu^I \hat \sigma_\mu$ and $\theta_\mu^J \hat \sigma_\mu$  can be summarized to one controlled-$R_z$.} \label{fig:control}
\end{figure*}

Now, we have two approaches to determine the unitary $e^{-i\Delta\beta \hat A}$. In the so-called state-specific approach, ${\bf a}$ is different for different $|\Phi_I\rangle$ (therefore we write ${\bf a}^{I}$ and $\hat A^I$ to indicate the state dependence). Similarly to QITE\cite{Motta20,Tsuchimochi22B}, we minimize the following function
\begin{align}
F^I({\bf a}^I) &=\left\|\sum_J d_{IJ} e^{-\Delta\beta (\hat H-E_J)}|\Phi_J^{(\ell)}\rangle - e^{-i\Delta\beta \hat A^I}|\Phi_I^{(\ell)}\rangle\right\|^2\label{eq:FI}
\end{align}
to the second-order of $\Delta\beta$ for each $I$. This results in the linear equation 
\begin{align}
{\bf M}^{I} {\bf a}^I + {\bf b}^I = {\bf 0}\label{eq:Ma+b=0}
\end{align}
with
\begin{align}
	M^I_{\mu\nu} & = 2 {\rm Re}\langle \Phi_I^{(\ell)} |  \hat \sigma_\mu \hat \sigma_\nu| \Phi_I^{(\ell)}\rangle\label{eq:AI}\\
	b^I_{\mu}& = {\rm Im} \langle \Phi_I^{(\ell)} | \left[\hat H, \hat \sigma_\mu\right]  | \Phi_I^{(\ell)}\rangle
	+  \frac{ 2}{\Delta\beta}\sum_{J} d_{JI} {\rm Im}\langle\Phi_I^{(\ell)}|\hat \sigma_\mu |\Phi_J^{(\ell)}\rangle\label{eq:bI}
\end{align}
We have provided a detailed derivation in the Supplementary Information.

In contrast, the state-averaged approach uses the same {\bf a} and $\hat A$ for all the states considered. This can be accomplished by solving 
\begin{align}
\sum_I {\bf M}^{I} {\bf a} + \sum_I {\bf b}^I = {\bf 0}.\label{eq:sumMa+b=0}
\end{align}

Several important considerations have to be made with respect to the above derivations.
Eqs.~(\ref{eq:AI}) and (\ref{eq:bI}) are essentially the same as those corresponding to QITE\cite{Tsuchimochi22B}, except that $b_\mu^I$ has the additional second term, which ensures the orthogonality of the model space. For the state-specific method, the model space is not exactly orthogonal, but it is almost so because of this term. Indeed, without the second term, Eq.~(\ref{eq:HcScE}) quickly becomes unsolvable because all elements of $S_{IJ}^{(\ell)}$ tend to become one (i.e., all states in the model space become the ground state). We note that $\frac{d_{IJ}}{\Delta\beta} \rightarrow 0$ for $I\ne J$ as $\Delta\beta \rightarrow 0$ and the diagonal term will not contribute because $\sigma_\mu$ is Hermitian and thus the expectation value is real; so the second term is stable.
The importance of the term is less pronounced in the state-averaged method. 

The state-averaged method would be preferred to the state-specific method because the model space $\{|\Phi_I^{(\ell)}\rangle\}$ is guaranteed to be orthogonal (i.e., $S^{(\ell)}_{IJ} = \delta_{IJ}$), and also because its quantum circuit is significantly simpler. However, despite the existence of a single unitary $e^{-i\Delta\beta \hat A}$ that correctly transforms all the states simultaneously to the desired states, it should be noted that the corresponding Hermitian $\hat A$ has to be quite complicated. In practice, because we truncate $\hat A$ after the single and double substitutions in Eq.~(\ref{eq:A}), the representability of the unitary is considerably limited, and therefore the performance of the state-averaged MSQITE may not be promising. This is quite similar to an issue recently reported by Ibe et al.\cite{Ibe22}, that the multistate contracted VQE, which minimizes the averaged energy of orthogonal states generated by the same unitary\cite{Parrish19}, experiences large errors for excited state calculations. Indeed, below, we will show that with the state-specific MSQITE a model space converges to almost the exact one using only single and double excitations in $\hat A$, whereas the accuracy of the state-averaged MSQITE is generally quite unsatisfactory and its errors in energy can be substantial especially when the number of states increases. 

\subsection*{Quantum Circuit for MSQITE}\label{sec:circuit}
The algorithmic difference between QITE and MSQITE is that the latter requires the estimation of quantities like $H_{IJ}^{(\ell)}$ for each pair $I,J$. Whereas the state-averaged MSQITE has a simple quantum circuit because all the states are evolved by the same unitary $e^{-i\Delta\beta \hat A}$, one needs the controlled gate for $e^{-i \Delta \beta \hat A({\bf a}^I)}$ for the state-specific approach. Figure \ref{fig:control} illustrates how we implement the latter using the Hadamard test. We prepare the state register and an ancilla qubit as $|\Phi_I^{(0)}\rangle$ and $|+\rangle$, which controls $U_{JI}$, $e^{-i\Delta\beta \hat A^I}$, and $e^{-i\Delta\beta \hat A^J}$. Here, $U_{JI}$ comprises simple gates to generate $|\Phi_J^{(0)}\rangle = \hat U_{JI} |\Phi_I^{(0)}\rangle$ initially. 
In practice, the unitary $e^{-i \Delta \beta \hat A({\bf a}^I)}$ is Trotter-decomposed as
\begin{align}
		e^{-i \Delta \beta \hat A({\bf a}^I)} \approx \prod_\mu e^{-i \theta_\mu^I \hat \sigma_\mu}
\end{align}
with
\begin{align}
	\theta_\mu^I = \Delta \beta a_\mu^{I}.
\end{align}
Since $\hat A^I$ and $\hat A^J$ only differ by the parameters ${\bf a}^I$ and ${\bf a}^J$ and share the same gate structure, it is convenient to order the controlled gates in an alternating manner as shown in Figure \ref{fig:control}(a), noting that the controlled-$e^{-i\theta_\mu^I \hat \sigma_\mu}$ and controlled-$e^{-i\theta_\nu^J \hat \sigma_\nu}$ always commute. Without the control qubit, each Pauli rotation is performed by using the standard procedure\cite{Nielsen-Chuang,Barkoutsos18, Romero19} as shown in Figure \ref{fig:control}(b), where (i) the qubits to be rotated are transformed to either of the $X, Y, Z$ basis by the corresponding single-qubit unitary gates (denoted by $R$), (ii) their parities are passed to the last qubit (denoted by the CNOT gate with a dotted line),  and (iii) the $R_z$ gate is applied followed by the Hermitian conjugate of (ii) and (i). Since the two adjacent controlled Pauli rotations carry out these unitary operations, the operations (i) and (ii) between them cancel out, and we can simplify the entire gate as depicted in Figure \ref{fig:control}(c).

Therefore, the additional complexity in the quantum circuit of the state-specific MSQITE arises from the two CNOT operations and one additional $R_z$ rotation. We consider this additional effort may not be a significant overhead cost compared with the circuit shown in Figure \ref{fig:control}(b).

For the diagonal terms $H_{II}^{(\ell)}$, they represent energy expectation values of $|\Phi_I^{(\ell)}\rangle$, and therefore their quantum circuits are identical to that of QITE, without any controlled-$e^{-i\Delta\beta \hat A}$ operations. Hence, for MSQITE with a model space comprising $n_{\rm states}$ states,  one needs to prepare $n_{\rm states}(n_{\rm states} -1)/2$ circuits at each $\ell$ to measure $H_{IJ}^{(\ell)}$ (noting that ${\bf H}^{(\ell)}$ is Hermitian), in addition to $n_{\rm states}$ circuits to perform the same measurements as QITE.

\begin{figure}[t!]
\includegraphics[width=0.45\textwidth]{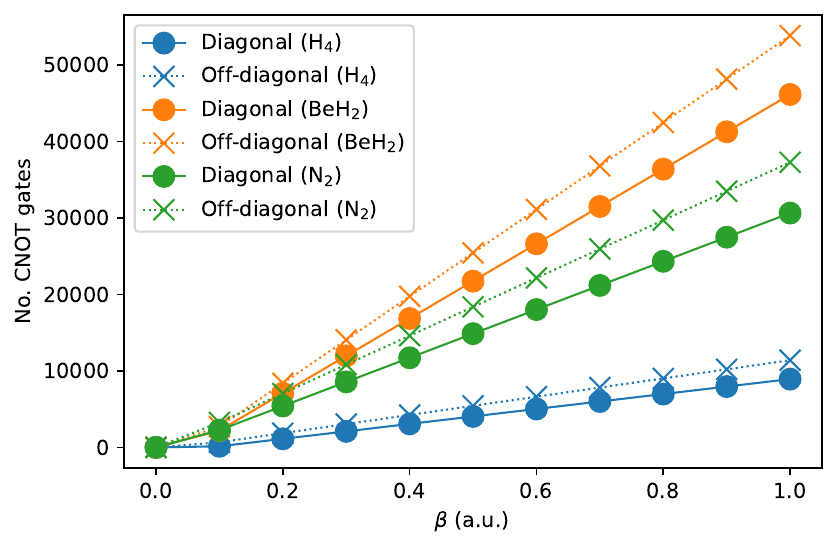}
\caption{Number of CNOT gates required to estimate the energy expectation value (diagonal) $\langle\Phi_I|\hat H|\Phi_I\rangle$ and coupling (off-diagonal) $\langle \Phi_I|\hat H|\Phi_J\rangle$ for each of noiseless simulations.}\label{fig:CNOT}
\end{figure}
To investigate the relative complexity of the quantum circuits required for the state-specific MSQITE, in Figure \ref{fig:CNOT}, we have depicted the number of CNOT gates for each noiseless simulation conducted in this study, as a function of imaginary time $\beta$. It is evident that the number of CNOT gates for the off-diagonal circuit is generally approximately 1.2 to 1.3 times greater than that for the diagonal circuit. This observation underscores the efficacy and effectiveness of our quantum circuit designed for the Hadamard test. Nevertheless, it should be mentioned that these quantum circuits are too deep if noise is of concern, and in practice we need to introduce appropriate simplifications, which will be discussed later.
\black

\subsection*{MS-QLanczos}\label{sec:MS-QLanczos}
Similar to many other Krylov methods\cite{Stair20, Seki21, Klymko22, Yoshikura23}, QLanczos forms and diagonalizes the effective Hamiltonian to generate a wave function as a linear combination of time-evolved states. Here, we can generalize QLanczos to the model space formalism, which we call MS-QLanczos. Let us consider to expand the Krylov model subspace as
\begin{align}
	\left\{  e^{-{\ell}\Delta\beta \hat H} |\Phi_I^{(0)}\rangle \; ; \; (\ell = 0,\cdots, n) \; ; \; (I = 1, \cdots, n_{\rm states}) \right\}\label{eq:MSQLanczos}
\end{align}
which comprises the basis for the effective Hamiltonian to be diagonalized. Here, we choose to use the normalized states $|\Phi^{(\ell)}_I\rangle$ to ease the derivation:
\begin{align}
	\left\{  |\Phi_I^{(\ell)}\rangle \; ; \; (\ell = 0,\cdots, n) \; ; \; (I = 1, \cdots, n_{\rm states}) \right\} \label{eq:MSQLanczos_subspace}
\end{align}
Note that it spans the same space as Eq.~(\ref{eq:MSQLanczos}). At an arbitrary time step $\ell\Delta\beta$, the $I$th quantum state is given by, 
\begin{align}
	|\Phi_I^{(\ell)}\rangle &= \sum_{J} d_{JI}^{(\ell -1)} e^{-\Delta \beta (\hat H-E_J^{(\ell -1)})} | \Phi_J^{(\ell - 1)}\rangle\nonumber\\
	&=  \sum_{J}\tilde d_{JI}^{(\ell -1)}  e^{-\Delta \beta (\hat H - E_0)} | \Phi_J^{(\ell - 1)}\rangle\label{eq:l_from_l-1}
\end{align}
where $E_0$ is some reference energy that is fixed throughout the imaginary time evolution (e.g., the average energy of the initial model space), and
\begin{align}
	&\Delta E_I^{(\ell)} = E_I^{(\ell)} - E_0 \\
	&\tilde d_{JI}^{(\ell)} = d_{JI}^{(\ell)} e^{\Delta\beta \Delta E_J^{(\ell)}} 
\end{align}
The global energy shift $E_0$ is introduced to ensure that the propagator is independent of both state and imaginary time, while avoiding the vanishing norm due to $e^{\Delta \beta E_0}$. 
 Using the relation (\ref{eq:l_from_l-1}) recursively, we find
\begin{align}
		|\Phi_I^{(\ell)}\rangle &= \sum_J\left(\tilde {\bf d}^{(\ell')} \cdots \tilde  {\bf d}^{(\ell -1)}\right)_{JI} e^{-(\ell -\ell') \Delta \beta (\hat H - E_0)} | \Phi_{J}^{(\ell')}\rangle
\end{align}
for arbitrary $\ell' < \ell$. 

Then, one can write the overlap matrix among the model space (\ref{eq:MSQLanczos_subspace}) as
\begin{align}
	{\mathscr S}_{I(\ell),J(\ell')} &\equiv\langle \Phi_I^{(\ell)} | \Phi_J^{(\ell')}\rangle \nonumber\\
&= \left[\left({\bf D}^{(\frac{\ell+\ell'}{2} \rightarrow \ell -1)}\right)^{ \top}  {\bf S}^{(\frac{\ell+\ell'}{2})} \left(  {\bf D}^{(\ell' \rightarrow \frac{\ell+\ell'}{2}-1)} \right)^{-1}\right]_{IJ}
\end{align}
where
\begin{align}
	 {\bf D}^{(\ell'\rightarrow \ell)} \equiv \tilde {\bf d}^{(\ell')} \tilde{\bf d}^{(\ell'+1)}\cdots\tilde{\bf d}^{(\ell)} \;\;\;\; (\ell > \ell')
\end{align}
Since we expect $n_{\rm states}$ to be small, the computational cost of ${\bf D}$ is negligible.
The MS-QLanczos Hamiltonian matrix elements are similarly derived as
\begin{align}
{\mathscr H}_{I(\ell),J(\ell')} &\equiv	\langle \Phi_I^{(\ell)} |\hat H| \Phi_J^{(\ell')}\rangle\nonumber\\
&= \left[\left({\bf D}^{(\frac{\ell+\ell'}{2} \rightarrow \ell -1)}\right)^{ \top} {\bf H}^{(\frac{\ell+\ell'}{2})} \left( {\bf D}^{(\ell' \rightarrow \frac{\ell+\ell'}{2}-1)} \right)^{-1}\right]_{IJ}.
\end{align}
and one simply solves the generalized eigenvalue problem using ${\bm {\mathscr H}}$ and ${\bm {\mathscr S}}$. 
Note that the derivation reduces to that of the modified single-state QLanczos\cite{Tsuchimochi22B} when $n_{\rm states} = 1$.

\subsection*{Illustrative noiseless simulations}

\begin{figure*}
\includegraphics[width=160mm]{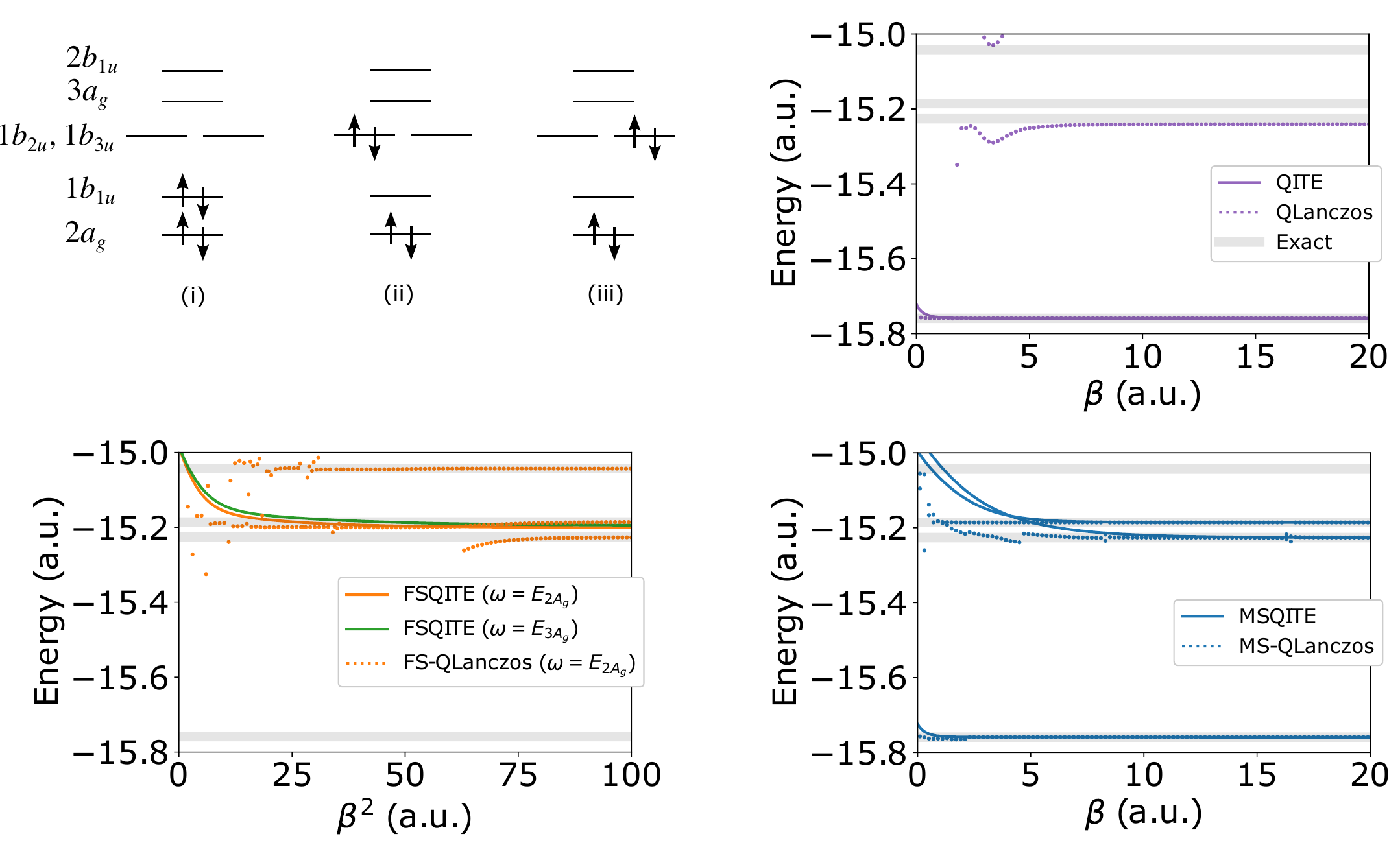}
\caption{Orbital diagrams and convergence for BeH$_2$ using different algorithms.}\label{fig:BeH2}
\end{figure*}

Here, we assess the performance of MSQITE and MS-QLanczos, using molecular systems without the impact of noise. For this reason, we use the unitary coupled-cluster generalized singles and doubles (UCCGSD) ansatz\cite{Barkoutsos18,Lee19}, which was found to be suitable for $\hat A$ in simulating molecules\cite{Tsuchimochi22B}.
Since the UCCGSD ansatz has the same number of parameters as the given Hamiltonian, the gate complexity for each time step in UCCGSD-based (MS)QITE scales similarly to the real time evolution. Therefore, the cost of UCCGSD-based MSQITE is expected to be comparable to algorithms that exploit real time evolution\cite{Stair20, Klymko22}.

We first consider the BeH$_2$ molecule at equilibrium (R$_{\rm e} = 1.334$ {\AA}). As the initial model space for MSQITE, we choose the following three configurations: the HF configuration, and the configurations where two electrons are promoted from the highest occupied orbital to $\pi$ orbitals, as listed in Fig.~\ref{fig:BeH2}. In the same figure, the performances of various methods for the ground and excited states are depicted. Because the ground state of the system is only weakly correlated, QITE and especially QLanczos quickly converge. Also shown in the figure is the results of FSQITE (using the exact target energy $E_{2A_g} = 15.2261$ Hartree) and its extension to QLanczos (FS-QLanczos). Although FSQITE and FS-QLanczos eventually converge to the exact states, their evolutions are rather slow. Moreover, it cannot determine the ground state because it is far from the target state. 

In contrast, clearly, MSQITE delivers remarkably fast convergence to the desired excited states when compared with FSQITE. Nevertheless, we note the convergence profile is state-dependent. For example, while the ground state $X A_g$ converged rapidly, the convergence of the second excited state $3A_g$ required approximately $\beta = 5$ {\it a.u.}, followed by the convergence of the first excited state $2A_g$ approximately $5\sim 6$ {\it a.u.} later. This difference is attributed to the fact that these excited states are strongly correlated. To see this, we have tabulated the coefficients of the exact eigenstates in Supplementary Table~S1. It is verified that the initial configurations (ii) $|000000110011\rangle$ and (iii) $|000011000011\rangle$ are the dominant ones for $3A_g$, each with a coefficient of about 0.5, but it also contains other dominant configurations such as $|000000111100\rangle$ (see the Methods section for our qubit mapping: here, two electrons are promoted from $2a_g$ to $1b_{2u}$ with respect to HF). Such additional configurations need to be generated by the (MS)QITE procedure, and the imaginary time evolution typically takes more steps if their coefficients are non-negligible. From Table~S1, it is seen that the  $2A_g$  state is even more strongly correlated than $3A_g$, resulting in slower convergence in MSQITE. 

We can expect a better performance of MSQITE if these additional configurations are included in the initial model space; however, of course, such detailed information may not be accessible {\it a priori}. Instead, MS-QLanczos can automatically detect and extract these states much earlier than MSQITE, as shown in Fig.~\ref{fig:BeH2}. In contrast to FS-QLanczos, MS-QLanczos was not able to obtain the $4A_g$ state. This is simply because we have truncated the Krylov vectors to avoid numerical instabilities. If such higher states are desired, one needs to add more states in the model space, and MSQITE (MS-QLanczos) can find the eigenstates in the energy order.

\begin{figure}
\includegraphics[width=80mm]{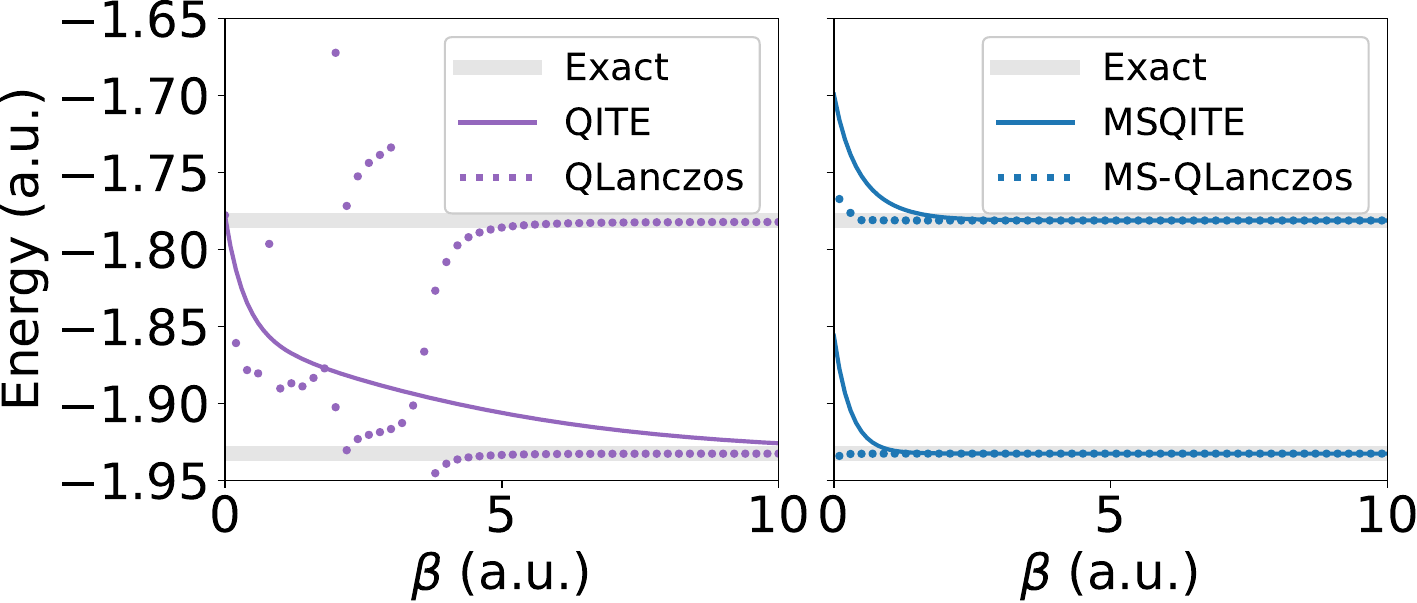}
\caption{Comparison between different algorithms for H$_4$.}\label{fig:H4}
\end{figure}

It should be noted that the MSQITE method should bring a certain advantage not only for excited states but also for strongly correlated ground states, because the model space by definition can naturally provide multi-configuration states. To observe this advantage, we take the square H$_4$ molecule with a bond length of 1 {\AA} as an example. As shown in Fig.~\ref{fig:H4}, QITE and QLanczos take more than 10 and 4 {\it a.u.} in imaginary time, respectively, to reach the ground state within the 1 mHartree accuracy. The slow convergence of the former is ascribed to the strong correlation in H$_4$, which is a two-determinant system with $|00001111\rangle$ and $|00110011\rangle$. 

The energies of MSQITE with the model space comprised of $|00001111\rangle$ and $|00110011\rangle$ approaches the (near) exact energies very rapidly, within less than a few {\it a.u.} in imaginary time; MS-QLanczos convergence is even faster. We emphasize, however, that each of the resulting MSQITE basis states $|\Phi_I\rangle$ are {\it not} the exact eigenstates. They are rather states that have either $|00001111\rangle$ or $|00110011\rangle$ as the dominant configuration, but possess almost no component of the other configuration. Nevertheless, the model space is developed to the complete space during the MSQITE procedure, such that linear combinations of $\{|\Phi_I\rangle\}$ are the exact states, as described in the preceding section.

\begin{figure}
\includegraphics[width=75mm]{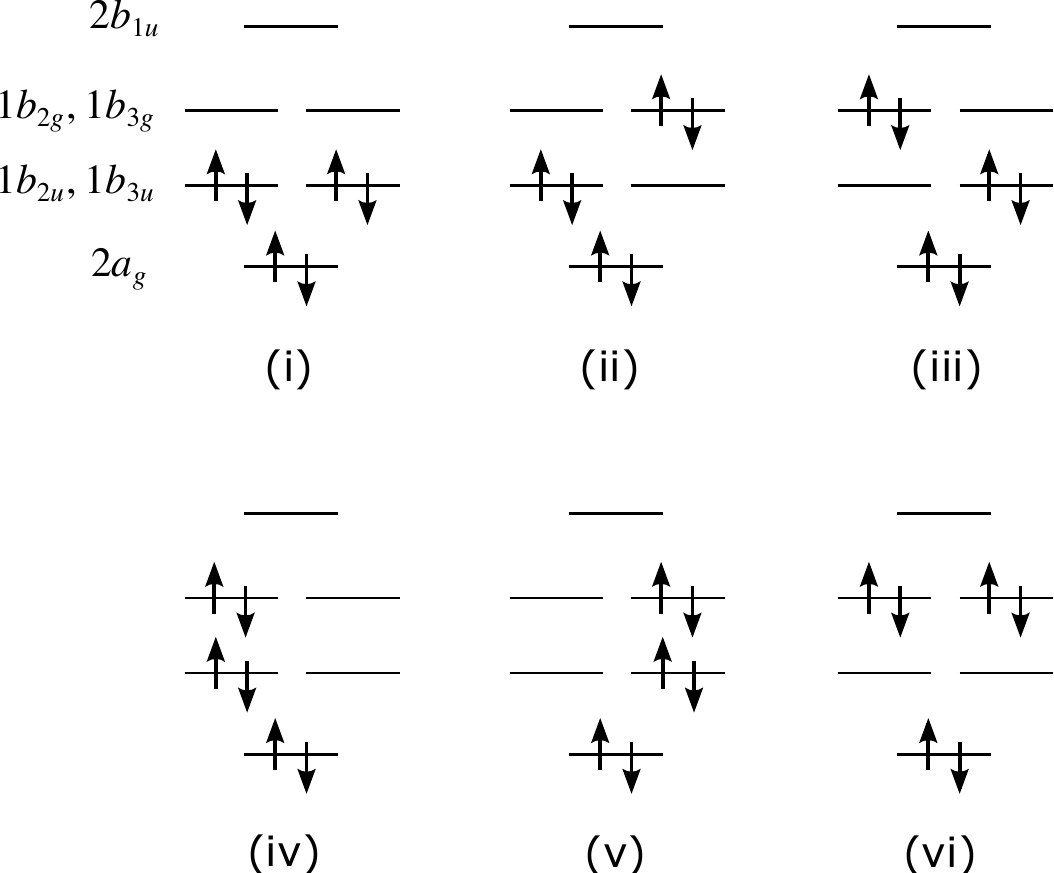}
\caption{Configurations used in MSQITE for N$_2$.}\label{fig:N2_diagram}
\end{figure}

\subsection*{Avoiding spin contamination with shifted propagator}\label{sec:spin-shift_results}
For a non-relativistic  molecular Hamiltonian, the exact wave function is an eigenstate of the number operator $\hat N$ and the spin operators $\hat S^2$ and $\hat S_z$. However, since each of the Pauli rotations applied in QITE does not necessarily commute with these symmetry operators, both the number of electrons and spin quantum numbers fluctuate during the evolution. Nevertheless, for the one-particle symmetry operators ($\hat N$ and $\hat S_z$), such fluctuations are moderate and do not affect the result in our numerical experiments. It is also relatively easy to constrict the quantum state to the fixed quantum numbers by using fermionic operators instead of Pauli operators, i.e., one can employ the parametrization of Eq.~(\ref{eq:A}) and treat linear combinations of Pauli operators\cite{Tsuchimochi22}. 

However, we found that the $\hat S^2$ symmetry is difficult to preserve, especially for excited states. Usually, the initial model space is prepared such that only the target spin states (e.g., singlets in the above cases) are included.  
However, due to the approximate nature of (MS)QITE, the model space often starts to leak into different spin symmetry spaces and finds higher spin states (e.g., triplets $s=1$ and quintets $s=2$) with lower energies than states with the desired spin, by virtue of imaginary time propagation. For variational simulations, one can use the projection operator\cite{Yen19, Tsuchimochi20, Tsuchimochi22}, but it is not straightforward to apply it in the framework of ITE. 

Such spin-contamination, and the resulting ``spin-collapse'', are  practical yet significant issues in MSQITE, as demonstrated below. In essence, when spin-collapse occurs, it becomes necessary to increase the number of states in a model space, $n_{\rm states}$, and rerun the calculation to obtain the true target states with low spins. However, it is important to emphasize that states with higher spin ($s$) than the target spin (i.e., singlet) can be obtained much more efficiently through a separate calculation, in which the initial states are prepared with $s = m_s = (N_\alpha-N_\beta)/2$, where $N_\alpha$ and $N_\beta$ are the numbers of $\alpha$ and $\beta$ electrons, respectively. Such simulations preclude the convergence of low-spin states (the fluctuation of $\hat S_z$ and hence $m_s$ is negligible in MSQITE). In fact, this protocol is a de facto standard frequently employed in quantum chemical calculations to track different spin states.

As an example, here we consider N$_2$ at a stretched bond distance of 1.6 {\AA}. We employed three configurations (i), (ii), and (iii) given in Fig.~\ref{fig:N2_diagram} to make an initial model space, and aimed to obtain the lowest singlet states with the $A_g$ symmetry including the ground state. 

\begin{figure}
\includegraphics[width=85mm]{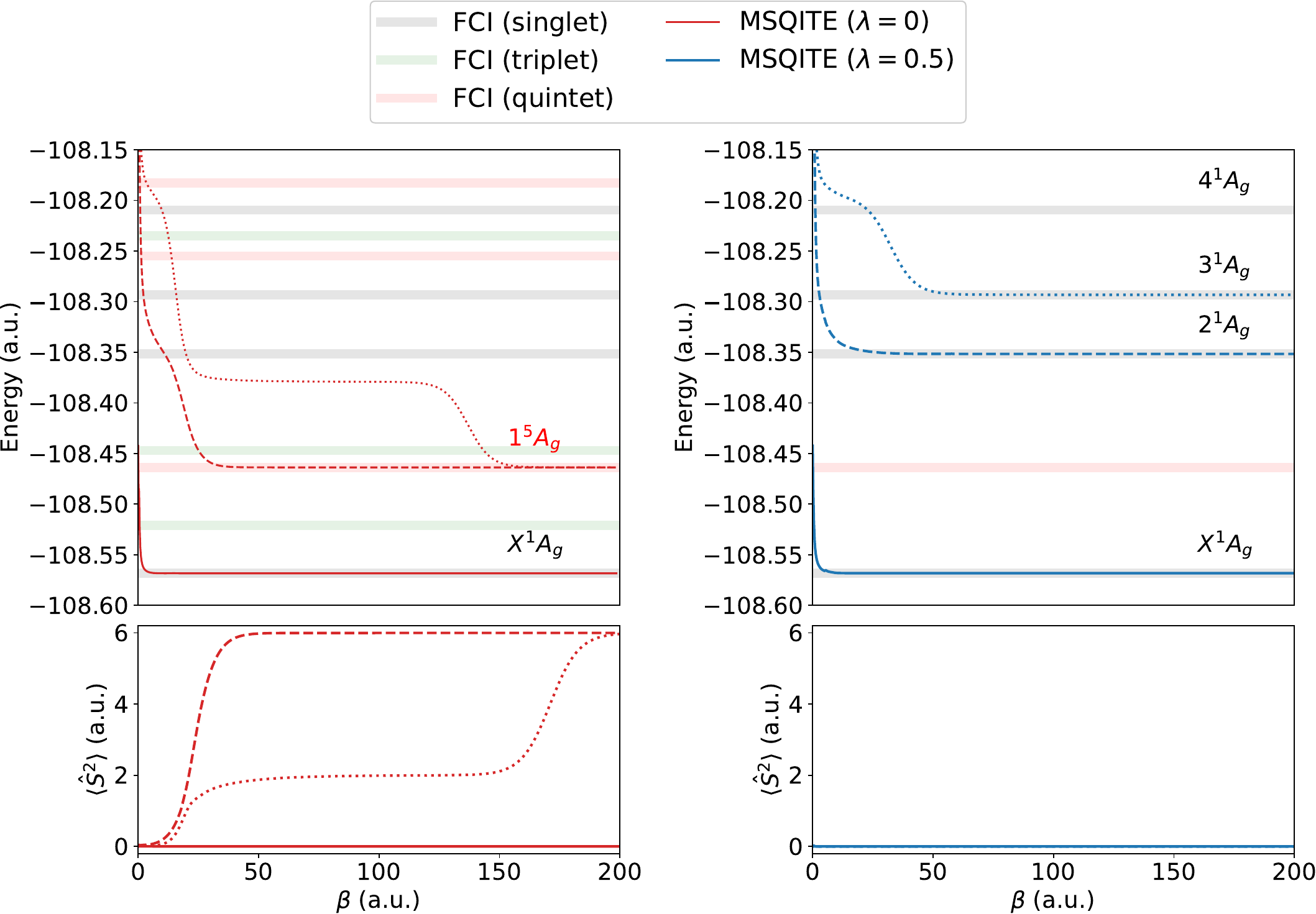}
\caption{Importance of spin-shift in MSQITE to remove spin-contamination (N$_2$ at a bond length of 1.6 {\AA}). Spin expectation values are plotted in the bottom figures.}\label{fig:spin-shift}
\end{figure}

The top left of Fig.~\ref{fig:spin-shift} plots the changes in energy of MSQITE state along with the exact energies of different spin symmetries (shown in different colors). Whereas the lowest state of MSQITE converges to the singlet ground state quickly, the two excited states suffer from slow convergence. More importantly, both converge to the quintet $1^5A_g$ state.  It is worth reiterating that  this lowest quintet state can be obtained simply by preparing an initial state with $m_s = 2$ in QITE, rather than adopting the more intricate MSQITE with $m_s = 0$. In the bottom left of the figure, we monitor the change in $\langle \hat S^2\rangle$, and the third state is trapped in some unphysical spin state with $\langle \hat S^2\rangle \approx 2$. 
The problem here is that, in general, the exact eigenstates are unknown and therefore one may get confused as if the MSQITE state achieved a stationary triplet state, as $\| {\bf b}_I \| \approx 0$ for $90 < \beta < 120$. However, this is an artifact of spin contamination.\footnote{This contaminated state was found to be a half-and-half mixture between $3^1A_g$ and $1^5A_g$: the exact energies are $E_{3^1A_g} = -108.293193$ and $E_{1^5A_g} = -108.463729$, and the energy expectation value of the state is trapped at about $-108.378 \approx (E_{3^1A_g} + E_{1^5A_g})/2$.}

Moreover, note that there exist several spin states (triplets and quintets) that have lower energies than singlet states as is clear from the figure. Hence, the convergence of MSQITE to these wrong spin states is highly likely. Of course, one could add more configurations in the model space to obtain higher singlet states; however, one can easily imagine that this approach is inefficient and is best avoided. We should also mention that, as the representability and flexibility of $\hat A$ increase, MSQITE would become even more prone to spin-contamination.

Hence, we introduce the spin-shift to the propagator,
\begin{align}
	e^{-\beta\hat H} \rightarrow e^{-\beta \left(\hat H + \lambda(\hat S^2 - s(s+1))\right)} \label{eq:spin-shift}
\end{align}
where $\lambda$ is an arbitrary positive number and $s$ is the designated spin quantum number. Because MSQITE is expected to transform the initial model space into the complete subspace, the use of the spin-shift should also be able to fix the spin at the same time. The trick here is that, while the target spin component with $s$ in the model space remains unaffected by the shifted propagator, the spin contaminants with $s' > s$ rapidly vanish. Note that we can always assume $s' \ge  s$ by appropriately constructing the initial model space (namely, we set $m_s = s$). Thereby, the model space will be projected to spin $s$.
We would like to emphasize that this effect shares similarities with the widely-used penalty function method in variational algorithms, which penalizes the energy associated with incorrect spin states\cite{Andrews91,McClean16, Yen19, Ryabinkin19}.

As $\lambda$ becomes large, the spin-projection acts more strongly; however, it could spoil the convergence of MSQITE because of the large Trotter error. In principle, it suffices to use $\lambda > E_{s'} - E_s$ where $E_{s'}$ is the energy of excited state with spin $s'$.

In the top right panel of Fig.~\ref{fig:spin-shift}, we show the results of MSQITE with the spin-shift using $\lambda = 0.5$. As expected, all the states nicely converge to the desired singlet states. Throughout all imaginary time, these states retain $\langle \hat S^2 \rangle =0$ approximately, and get rid of spin contamination appropriately. We notice that the third state of MSQITE initially approaches the $4^1A_g$ state instead of directly converging to the $3^1A_g$ state, and then starts to find the latter state as the lower state. However, this is not the weakness of the method; it is rather an indication of the ability of MSQITE to find the lowest states.

\subsection*{State-specific and state-averaged MSQITE}\label{sec:SS_vs_SA}
In the preceding section, we have discussed the advantages that MSQITE has to offer, focusing on the state-specific algorithm. As the state-averaged scheme is more attractive in terms of circuit complexity, we also carried out the state-averaged MSQITE to evaluate its accuracy. To properly evaluate the potential of the state-averaged MSQITE, we have performed noiseless simulations. 

\begin{figure}[b]
\includegraphics[width=83mm]{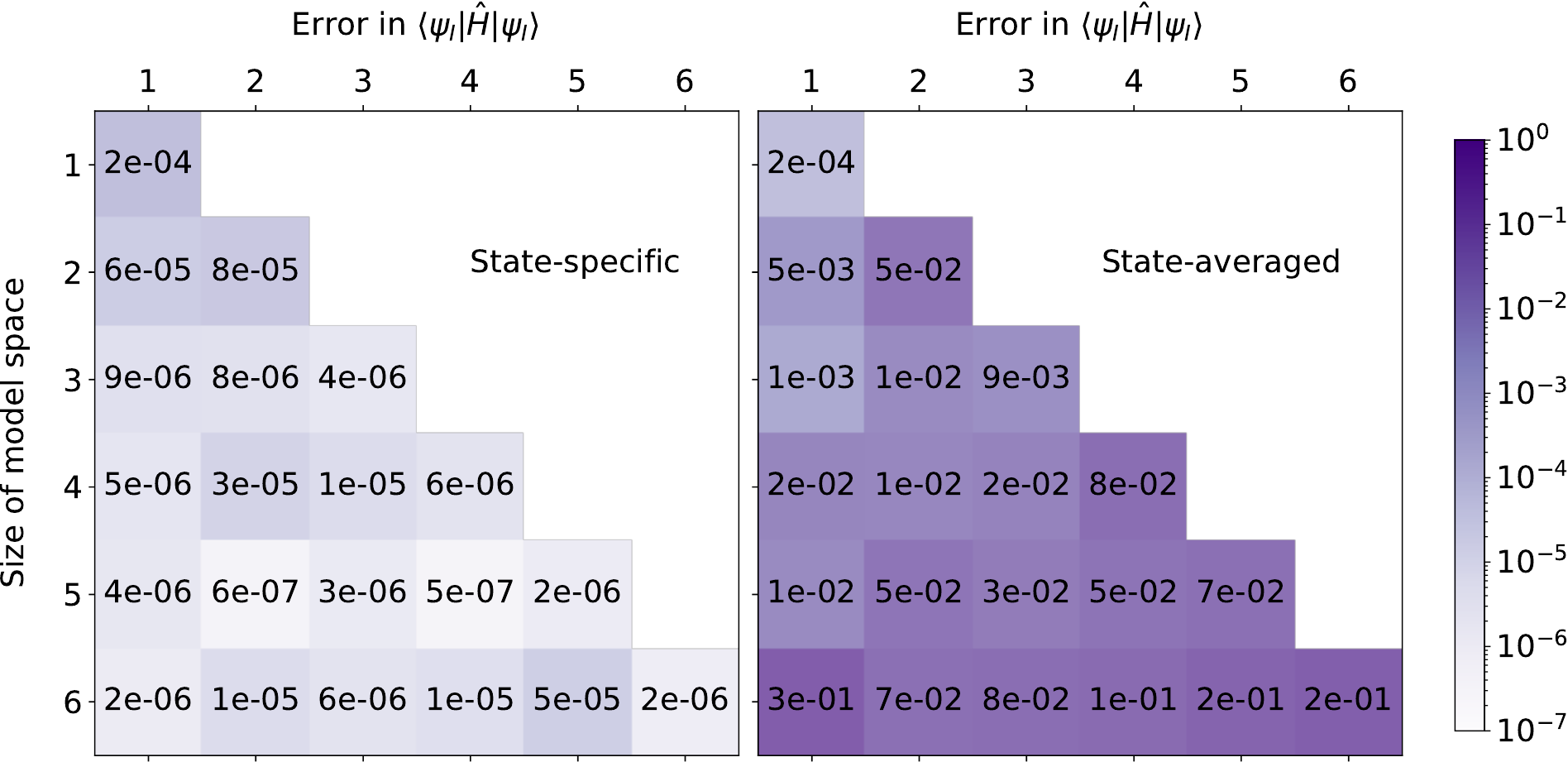}
\caption{Energy errors of the converged UCCGSD-based MSQITE states with the state-specific (left) and state-averaged (right) schemes for N$_2$ at equilibrium. States used for each MSQITE simulation are chosen from Fig.~\ref{fig:N2_diagram} in serial order.}\label{fig:N2_mat}
\end{figure}

Table~\ref{tb:SS_vs_SA} compares the final energies obtained at convergence of the state-specific and state-averaged MSQITE methods. In addition to H$_4$ and BeH$_2$, N$_2$ at equilibrium (a bond distance of 1.098 {\AA}) was tested with two configurations (i) and (ii) in Fig.~\ref{fig:N2_diagram},  as an initial model space. Whereas the state-specific MSQITE yields quite accurate energies independent of systems, the state-averaged MSQITE results become significantly inaccurate for larger systems. Its accuracy is satisfactory for H$_4$ but deteriorates for N$_2$ with an error of 47 mHartree for the $2A_g$ state. In general, increasing the model space tends to result in larger errors in energy, as shown in Fig.~\ref{fig:N2_mat}. 

\begin{table} 
\caption{Exact energy and error of MSQITE  for each system (in Hartree).}\label{tb:SS_vs_SA}
\begin{tabular}{lrrr}
\hline\hline
System &Exact\;\;\;\;\;\; & State-specific & State-Averaged \\
\hline
H$_4$ $XA_g$ 	&	-1.932 645	&	$<1\times 10^{-8}$	&	$2\times 10^{-7}$	\\
H$_4$ $2A_g$ 	&	-1.781 254	&	$<1\times 10^{-8}$	&	$2\times 10^{-7}$	\\
BeH$_2$ $XA_g$ 	&	-15.759 026	&	$<1\times 10^{-8}$	&	$2\times 10^{-4}$	\\
BeH$_2$ $2A_g$	&	-15.226 336	&	$8\times 10^{-8}$	&	$4\times 10^{-4}$	\\
BeH$_2$ $3A_g$	&	-15.185 771	&	$<1\times 10^{-8}$	&	$2\times 10^{-4}$	\\
N$_2$ $XA_g$	&	-108.669 173	&	$6\times 10^{-5}$	&	$5\times 10^{-3}$	\\
N$_2$ $2A_g$	&	-107.968 085	&	$8\times 10^{-5}$   &	$5\times 10^{-2}$	\\
\hline\hline
\end{tabular}	
\end{table}

\begin{figure*}[t!]
\includegraphics[width=0.99\textwidth]{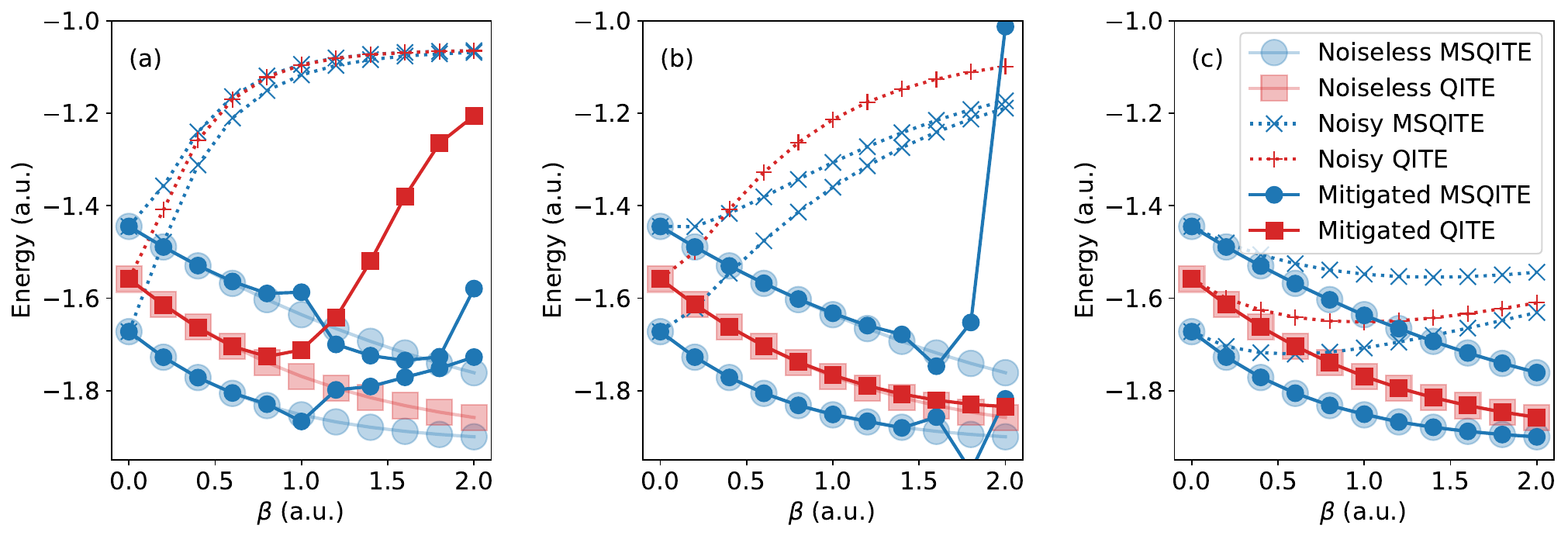}
\caption{Noisy QITE and MSQITE for H$_4$ using depolarizing error. The error rate for one-qubit gates $p_1$ is set to $0.1$ times the error rate for two-qubit gates $p_2$. (a) $p_2=10^{-2}$ (b) $p_2 = 5\times 10^{-3}$, and (c) $p_2 = 10^{-3}$.}\label{fig:Noise}
\end{figure*}

Another prominent example of the failure of the state-averaged MSQITE is the N$_2$ molecule with two $\pi$ orbitals and two $\pi^*$ orbitals and four electrons (comprising an eight qubit system). With a model space comprising six configurations --- HF and all five pair-excited configurations derived from it (those listed in Fig.~\ref{fig:N2_diagram})--- the UCCGSD-based state-averaged MSQITE methods immediately converge at $\beta=0$, because $\sum_I {\bf b}^I = {\bf 0}$ by symmetry. Note that this convergence does not indicate, of course, ${\bf b}^I = {\bf 0}$ for each state; in fact, the state-specific MSQITE performs quite well, yielding very accurate energies. It is worth noting that $\sum_I {\bf b}^I$ is equivalent to the averaged energy derivative that appear in the VQE-based state-averaged UCCGSD method\cite{Yalouz21}; indeed, we applied the method to this system and found that it suffers from the same problem and no optimization of parameters was carried out. Overall, this strongly implies the limitation of other state-averaged methods for general systems\cite{Ibe22, Parrish19}.

It should be clear that this ill-behavior of the state-averaged MSQITE does {\it not} necessarily imply a possible theoretical flaw in our derivation. The failure is rather ascribed to the limitation of the form of $\hat A$ that we employed, i.e., single and double substitutions. In other words, it is unlikely the same UCCGSD amplitudes can evolve any arbitrary states to the desired ones all at once through $e^{-i\Delta\beta\hat A}$, even qualitatively. That being said, with triples (T) and quadruples (Q) included, we can rigorously obtain the exact eigenstates by definition: such UCCGSDTQ ansatz is complete for a four-electron system. For this particular case, the UCCGSDT-based state-averaged MSQITE already delivers almost the exact result (with less than $10^{-12}$ Hartree error). 

Overall, therefore, we are led to conclude that the state-averaged MSQITE does not seem practical because one needs way more Pauli operators from higher rank excitations than double excitations, to achieve a satisfactory accuracy, and this will become quickly infeasible with the increase in number of electrons.

\subsection*{Noise simulations}
To investigate the effect of quantum noise in state-specific MSQITE, here we conducted noise simulations on squared H$_4$ with each side being $2$ \AA. To simplify the quantum circuits, we have used several techniques (refer to the the METHODS section for details). Figure \ref{fig:Noise} showcases both noiseless and noisy results for QITE and MSQITE, using different error rates for depolarizing noise. As is evident from the figure, both noisy QITE and noisy MSQITE experience significant error accumulation as imaginary time extends. For a relatively conservative error rate for each CNOT, $p_2=10^{-2}$, the energies for both noisy QITE and noisy MSQITE become exceed that of HF already at $\beta = 0.2$ {\it a.u.} (see Figure \ref{fig:Noise}(a)). 

\begin{figure}[b!]
\includegraphics[width=0.45\textwidth]{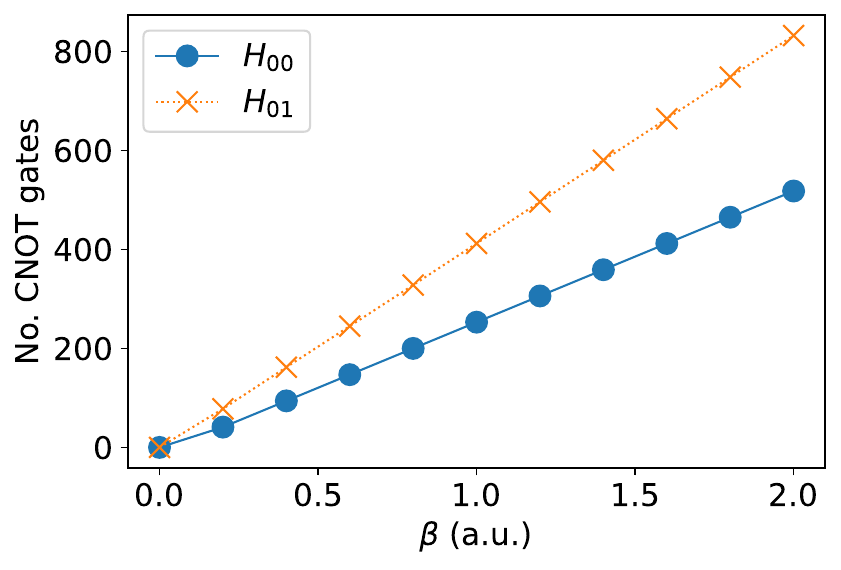}
\caption{Number of CNOT gates required to estimate the energy expectation value (diagonal) $\langle\Phi_I|\hat H|\Phi_I\rangle$ and coupling (off-diagonal) $\langle \Phi_I|\hat H|\Phi_J\rangle$ in the noisy simulations of H$_4$.}\label{fig:CNOT_H4}
\end{figure}
Such large errors arise from the fact that the circuit depth increases linearly with the imaginary-time step in these methods. Figure \ref{fig:CNOT_H4} plots the number of CNOT gates required to estimate the diagonal and off-diagonal matrix elements of the effective Hamiltonian, $H_{00}$ and $H_{01}$, respectively, as a function of $\beta$. We will omit the results for $H_{11}$ because they are almost the same as the results of $H_{00}$. As discussed earlier, the quantum circuit for $H_{01}$ is not significantly more complicated compared to that for $H_{00}$. Nevertheless, the number of CNOT gates exceeds 400 at $\beta = 1$ {\it a.u.} (corresponding to five imaginary-time steps, as we used $\Delta \beta = 0.2$ {\it a.u.} for these simulations), which introduces tremendous errors even with a small error rate of $p_2=10^{-3}$ for each CNOT gate (Figure \ref{fig:Noise}(c)). Especially, as the noise accumulates, the expectation values of off-diagonal matrix elements tend to zero, leaving almost no advantage in performing MSQITE over QITE.

\begin{figure}[t!]
\includegraphics[width=0.45\textwidth]{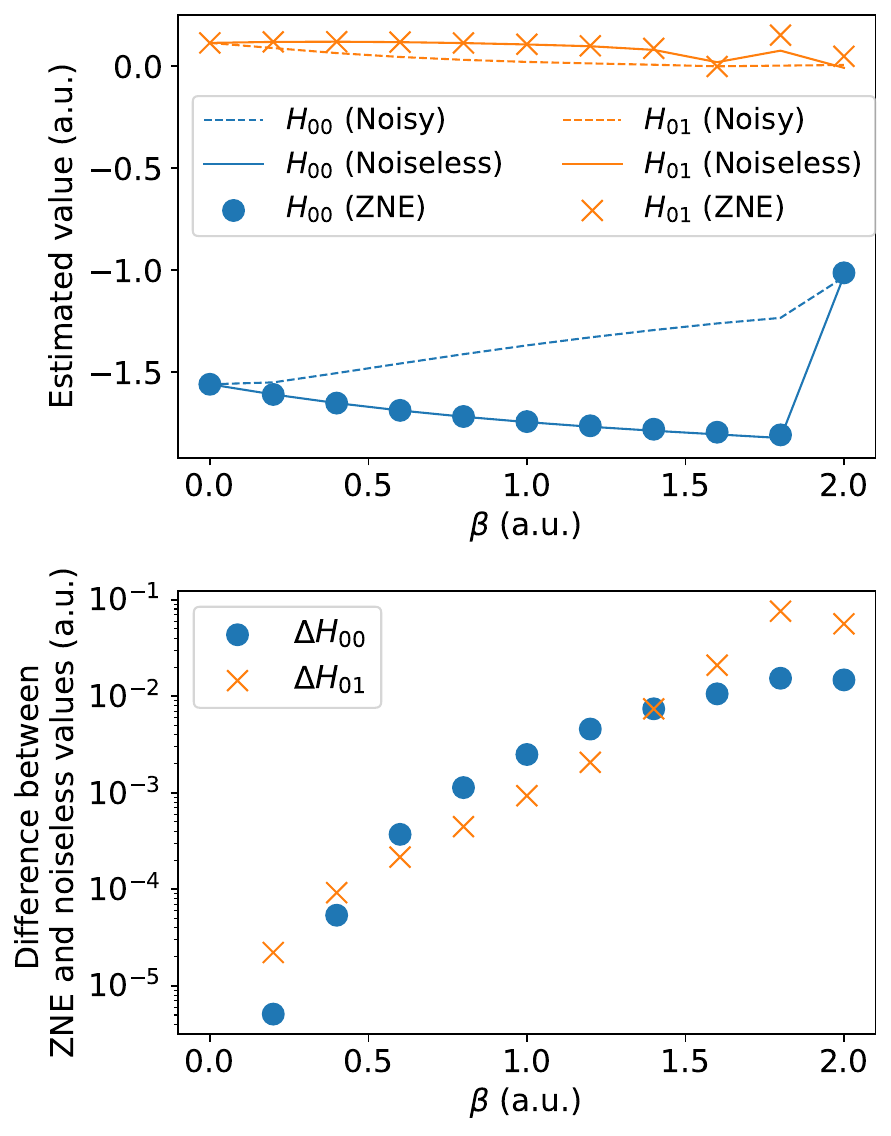}
\caption{(Top) Noisy, noiseless, and extrapolated values of the diagonal and off-diagonal matrix elements $H_{00}$ and $H_{01}$ for $p_2 = 5\times 10^{-3}$. (Bottom) Difference between extrapolated and noiseless values.}\label{fig:Noise_2}
\end{figure}

Therefore, it is essential to mitigate errors with NISQ computers\cite{Kandala19, Kim23}, and we have employed the simple protocol of zero-noise-extrapolation (ZNE)\cite{Li17, Temme17, Endo18}. The error-mitigated QITE and MSQITE results are also presented in Figure \ref{fig:Noise} as filled squares and circles, respectively. Clearly, ZNE  significantly improves the noisy results, successfully approximating the noiseless energies at short $\beta$ values. Interestingly, we observe that error-mitigated MSQITE is almost always as stable as mitigated QITE, despite the slightly longer circuit required for $H_{01}$. 

This result can be rationalized by recognizing that the off-diagonal elements generally have a much smaller absolute value than the diagonal ones, serving as a correction to the QITE energies. As a result, errors in the off-diagonal elements have a less impact on the accuracy of the final energies of MSQITE, compared to errors in the diagonal elements.

In order to examine this result further, we plotted the noisy values ($p_2 = 5\times 10^{-3}$) of $H_{00}$ and $H_{01}$ along side the extrapolated values in the top panel of Figure \ref{fig:Noise_2}. The ``ideal'' noiseless estimates, generated using the same quantum circuits and parameters ${\bf a}$, are also shown. We observe that the noisy $H_{01}$ converges to 0, as expected.  In most instances, ZNE successfully approximates the noiseless values of $H_{00}$ and $H_{01}$ with notable accuracy. The bottom panel of Figure \ref{fig:Noise_2} illustrates the differences between the ZNE and noiseless values, namely, $\Delta H_{IJ} = \left|H_{IJ}({\rm ZNE}) - H_{IJ}({\rm Noiseless})\right|$ for $IJ \in (00, 01)$, which serve as a metric for the reliability of ZNE. Remarkably, both $\Delta H_{00}$ and $\Delta H_{01}$ exhibit similar precision, even though $H_{01}$ is presumed to be more susceptible to noise  due to its higher CNOT gate count. Therefore, the errors introduced in the effective Hamiltonian are relatively balanced between diagonal and off-diagonal elements, yielding an overall noise impact on the MSQITE energies similar to that on the QITE energies.

In conclusion, MSQITE continues to offer its advantages over QITE even on NISQ computers, as long as suitable error-mitigation techniques are utilized to minimize the effects of noise.

\section*{Discussion}\label{sec:conclusion}
In this work, we introduced a model space into QITE to enable excited state simulations. The orthogonality condition was retained using the L\"owdin symmetric orthonormalization, which minimizes the state change during time steps and thus is suitable for the short-time unitary approximation of the non-unitary imaginary time propagation. MSQITE was shown to be a promising route to obtaining both ground and excited states, and its extension to QLanczos allowed for further acceleration in obtaining approximate eigenstates. 

This study also proposed the spin-shift in the propagator. Because excited states frequently suffer from spin-contamination, it is necessary to remove the irrelevant spin configurations from the MSQITE simulation. We have shown that the proposed spin-shift approach achieves this feat by projecting out the desired spin symmetry through ITE.

Based on the results obtained in this work, we conclude that the state-averaged MSQITE is likely to necessitate substantially complicated $\hat A$ to appropriately evolve all the states considered in a model space, compared to the state-specific scheme.  Namely, the former method requires fermionic excitations beyond double excitations in $\hat A$ for $e^{-i\Delta\beta \hat A}$ to achieve reasonable accuracy  in ideal, noiseless computations; this is deemed unappealing because of the increasing number of Pauli operators that need to be included. It should be pointed out that the recently proposed state-averaged orbital-optimized VQE\cite{Yalouz21} shares the same difficulty because it uses the same unitary for multiple orthonormal states as in the state-averaged MSQITE; thus, the scalability of the method with the increase in number of electrons and states remains to be an open question. In contrast, the state-specific MSQITE is potentially more promising than the state-averaged one, requiring only single and double excitations (i.e., UCCGSD) to achieve high accuracy. 

We have also demonstrated that the effectiveness of state-specific MSQITE compared to QITE remains unaffected by quantum noise when appropriate error mitigation techniques are applied. This resilience can be attributed to two main factors. First, the quantum circuit for state-specific MSQITE was designed to minimize the CNOT gate counts in the estimation of off-diagonal elements. Second, both the diagonal and off-diagonal elements display a comparative magnitude of errors.
 Overall, we found that the impact of quantum noise on MSQITE is similar to that on QITE.
 
 Finally, many of the ideas developed in this work are versatile, and we expect that they can be applied to fields beyond quantum chemistry, such as nuclear physics. Moreover, MSQITE can be extended to combine with adaptive algorithms\cite{Gomes20, Yeter-Aydeniz20} and variational algorithms\cite{McArdle19, Gomes21}. This integration holds potential for new synergies that can reduce the circuit depth. We are currently working along these directions.

\section*{Methods}

\subsection*{Orthonormalization of model space}\label{app:lowdin}
In the Results section, the transformation matrix {\bf d} was introduced in MSQITE to preserve the orthonormality of the model space after a short-time propagation. We chose the L\"owdin symmetric orthonormalization for this purpose. First, we form the overlap matrix of the target imaginary time evolved model space,
\begin{align}
\tilde S_{IJ}^{(\ell)} &= \langle \Phi_I^{(\ell)}|e^{-\Delta\beta (\hat H - E_I)} e^{-\Delta\beta (\hat H - E_J)}|\Phi_J^{(\ell)}\rangle \nonumber
\\&	= S_{IJ}^{(\ell)}   - 2 \Delta \beta \left( H_{IJ}^{(\ell)} -\frac{1}{2}(E_I + E_J) S^{(\ell)}_{IJ}\right) + O(\Delta\beta^2) \label{eq:Stilde}
	\end{align}
which is truncated after the first-order of $\Delta\beta$ to obtain the approximate overlap. 
Note that here we assume the model space is orthonormal $S^{(\ell)}_{IJ}=\delta_{IJ}$; however, even if this assumption is not satisfied, one can still find such a basis and the argument does not lose generality, see the Supplementary Information. 
Diagonalizing $\tilde {\bf S}$ gives
\begin{align}
	\tilde {\bf S}{\bf U} = {\bf U}\tilde {\bf s}
\end{align}
where {\bf s} is the diagonal matrix with the eigenvalues and {\bf U} the eigenvectors. The {\bf d} matrix from the L\"owdin symmetric orthonormalization is then uniquely obtained as
\begin{align}
	{\bf d} = {\bf U} \tilde {\bf s}^{-1/2}{\bf U}^\dag.
\end{align}

We note that, instead of the above {\bf d}, it would be also tempting to employ the transformation that diagonalizes $\langle \Phi_I^{(\ell)}| e^{-\Delta\beta (\hat H - E_I)}\hat H e^{-\Delta\beta (\hat H-E_J)}|\Phi_J^{(\ell)}\rangle$, such that $\lim_{\ell \rightarrow \infty}|\Phi_I^{(\ell)}\rangle$ is the exact ground or excited state,  $|\psi_I\rangle$. However, it is easily seen that the unitary matrix obtained from the diagonalization of such an effective Hamiltonian matrix is inadequate because it can flip the signs and even the ordering of the states, and thus Eq.~(\ref{eq:MSQITE}) cannot be a valid approximation. 

Following Blunt et al.\cite{Blunt15}, one may perform the Gram-Schmidt orthogonalization to define {\bf d}. However, the Gram-Schmidt orthogonalization is not unique about the order of orthogonalization steps and also leads to a biased update of $\{|\Phi_I\rangle\}$. Importantly, the propagation of the first state $|\Phi_0\rangle$ will remain unaffected by the presence of other states $|\Phi_I\rangle (I>0)$. Therefore, it will naturally become the exact ground state at $\beta \rightarrow \infty$. It is highly desirable that $|\Phi_0\rangle$ is initially chosen to be the closest to the ground state among all the states in the model space at $\beta =0$. Otherwise, the model space would experience large reorganization, which the short-time unitary evolution of Eq.~(\ref{eq:MSQITE}) would find difficult to express. This requirement may be easily satisfied for the ground state (i.e., HF may be the most reasonable starting point). However, for excited states, the appropriate ordering is generally unknown.

\subsection*{Simulation details}
MSQITE and MS-QLanczos were implemented in our Python-based emulator package, {\sc Quket}\cite{quket}, which compiles other useful libraries such as {\sc OpenFermion}\cite{openfermion}, {\sc PySCF}\cite{pyscf}, and {\sc Qulacs}\cite{qulacs}, to perform quantum simulations. In all simulations, we used the STO-6G basis set and HF orbitals. The Jordan-Wigner transformation was employed to map the fermion operators to the qubit representation, such that $\alpha$ and $\beta$ spin orbitals were aligned alternately with the rightmost qubit represents the lowest energy $\alpha$ spin orbital.  $\Delta\beta$ was set to 0.1 {\it a.u.} for QITE and MSQITE, and 0.05 {\it a.u.} for FSQITE, and the UCCGSD ansatz\cite{Barkoutsos18, Lee19, Tsuchimochi22B} was employed for the form of $\hat A$, except for the noise simulations, for which the computational details are provided below. To ensure stable solutions to Eqs.(\ref{eq:Ma+b=0}) and (\ref{eq:sumMa+b=0}), we applied a truncation scheme to the singular values of the {\bf M} matrix. Specifically, we retained singular values that were greater than $10^{-7}$ times the largest singular value, as implemented in {\tt scipy.linalg.lstsq}\cite{scipy}. The Be 1$s$ orbital and the N 1$s$ and 2$s$ orbitals were not considered in the simulations. For H$_4$, the initial HF calculation was performed with the $C_{2h}$ symmetry instead of $D_{4h}$, to relax the orbitals.

To perform noisy QITE simulations, we used Qiskit\cite{qiskit}, and the quantum noise was modeled by depolarizing error. The error rate for two-qubit gates was set to $p_2 = 10^{-2}, 5\times10^{-2},$ and $10^{-3}$, and that for one-qubit gates was set to $p_1 = p_2\times 0.1$. To mitigate the error, we used the zero-noise-extrapolation technique with exponential fitting\cite{Endo18}, only for two-qubit (CNOT) gates. For this simulation, the Hamiltonian of H$_4$ is transformed to the reduced representation by using the tapering-off technique\cite{Bravyi17,Setia20}. This resulted in the following four-qubit-Hamiltonian:
\begin{align}
\hat H=& h_0
+ h_1 Z_2 Z_3
+ h_2 Z_0 Z_2
+ h_2 Z_1 Z_2
+ h_3 Z_0 Z_1 Z_2 Z_3\nonumber\\
&+ h_4 Z_0 Z_1
+ h_5 (Z_0 Z_3  +  Z_1 Z_3)
+ h_6 (Z_0 Z_1 Z_2 + Z_0 Z_1 Z_3)\nonumber\\
&+ h_7 (-X_1 Y_2 Y_3+Y_1 Y_2 X_3-X_0 Y_2 Y_3+Y_0 Y_2 X_3)\nonumber\\
&+ h_8 (-Y_0 Z_1 Y_2 Z_3 + Y_0 Z_1 Z_2 Y_3 + Y_0 Y_2 + X_0 Z_1 Z_2 X_3 
\nonumber\\
&- Z_0 X_1 Z_2 X_3 +Z_0 Y_1 Y_2 Z_3 - Z_0 Y_1 Z_2 Y_3 - Y_1 Y_2)\nonumber\\
&+ h_9(-X_0 Z_1 Z_2 + X_0 Z_2 Z_3 - Z_0 X_1 Z_2 + X_1 Z_2 Z_3)\nonumber\\
&+ h_{10}(X_0 X_1 Z_2 Z_3 + Y_0 Y_1 Z_2 Z_3)\nonumber\\
&+ h_{11}(X_0 Y_1 Y_2          - Y_0 X_1 Y_2 + Z_0 Z_2 X_3 - Z_1 Z_2 X_3)
\nonumber\\
&+ h_{12}(Z_0 Z_1 Y_2 Y_3  - Y_2 Y_3)
\nonumber\\
&+ h_{13}(Z_0 + Z_1 + Z_0 Z_2 Z_3 + Z_1 Z_2 Z_3)
\nonumber\\
&+ h_{14}(-Z_2-Z_3)\
\nonumber\\
&+ h_{15}(-Y_0 X_1 Y_3 + X_0 Y_1 Y_3 + Z_0 X_2 Z_3 - Z_1 X_2 Z_3)
\end{align}
with
\begin{subequations}
    \begin{align}
h_{0 }&=-1.0613356242517709\\
h_{1 }&=0.3752318182963852\\
h_{2 }&= 0.3736748877137335\\
h_{3 }&= 0.3722054115496151\\
h_{4 }&= 0.26033869205932514\\
h_{5 }&= 0.22299864958557725\\
h_{6 }&= 0.09825586237928423\\
h_{7}&=0.07901175885991207\\
h_{8}&=0.0746839486241239\\
h_{9}&=0.07166447926823467\\
h_{10}&=0.05690174793752179\\
h_{11}&=0.05592385851947026\\
h_{12}&=0.05496497155276822\\
h_{13}&=0.03491578410706995\\
h_{14}&=0.021711508654056723\\
h_{15}&=0.018760090104653643
    \end{align}
\end{subequations}
whose first two lowest eigenvalues are $-1.91552763$ and $-1.87493645$.
We also simplified the quantum circuit by using a reduced form of UCCGD (neglecting single substitutes) where only one of eight pauli tensors arising from a double excitation is used. 
The resulting form of the ansatz is comprised of 14 unitaries,
\begin{align}
    e^{-i\Delta \beta \hat A} = \prod_{i=0}^{13} e^{-i\Delta \beta a_i\hat P_i}
\end{align}
with
\begin{subequations}
\begin{align}
\hat P_0 &= X_1 X_2 Y_3 \\
\hat P_1 &= X_0 X_2 Y_3\\
\hat P_2 &= Z_1 Z_2 Y_3 \\
\hat P_3 &= X_0 Y_1 X_2 \\
\hat P_4 & = X_2 Y_3\\
\hat P_5 &= X_0 Y_3\\
\hat P_6 &= X_1 Y_2\\
\hat P_7 &= Y_0 Z_1 X_2 Z_3\\
\hat P_8 &= Z_0 Y_1 Z_2 X_3\\
\hat P_9 &= Y_0 X_1 X_3\\
\hat P_{10} & = Z_0 Y_2 Z_3\\
\hat P_{11} & =X_0 Y_1 \\
\hat P_{12} &= Y_1 \\
\hat P_{13} &= Y_0
\end{align}
\end{subequations}
where we have used $\Delta \beta = 0.2$ {\it a.u.}.
 \black
We used the stabilization procedure for QLanczos as presented in Ref.~[\onlinecite{Tsuchimochi22B}] to describe excited states. However, for MS-QLanczos, the numerical instability arising from the linear dependence in the Krylov subspace becomes even more challenging compared with the standard QLanczos. Hence,  we adopted the same procedure as Ref.[\onlinecite{Motta20}], i.e., we use Krylov vectors that satisfy ${\mathscr S}_{\ell \ell'} < 0.99$ to alleviate the linear dependence. However, we have made the following modifications: the selection of Krylov vectors is performed backwards (i.e., starting from the current time instead of from the initial time) in order to ensure the latest states are always included in the basis, and the number of states included in the subspace is limited to 5. The selection is based on the assumption that excessively old time states do not play an important role but only cause numerical instabilities.

\section*{Data Availability}
The data that support the findings of this study are available from the corresponding author upon reasonable request.

\section*{Code Availability}
The code that is used to produce the data presented in this study is available from the authors upon reasonable request.

\section*{Acknowledgements}
This work was supported by JST, PRESTO (Grant Number JPMJPR2016), Japan and  by JSPS KAKENHI (Grant Number JP20K15231). We are grateful for the computational resources provided by ECCSE, Kobe University.
\section*{Competing interests}
The authors declare no competing financial or non-financial interests.

\section*{Author contributions}
T.T. conceived the idea and wrote the paper. Y.R., S.C.T., and T.T. implemented the algorithms and performed numerical simulations. T.T., Y.R., S.C.T., and S.L.T. all participated in discussions that developed the theory and shaped the project.

\end{document}